\documentstyle[fleqn,12pt]{article}

\def\thalf{{\textstyle{\frac{1}{2}}}}
\def\tquar{{\textstyle{\frac{1}{4}}}}
\def\oneth{{\textstyle{\frac{1}{3}}}}
\def\twoth{{\textstyle{\frac{2}{3}}}}
\def\pj{\hspace{-.26cm}}
\begin{document}
\title{Role of Hyperon Negative Energy Sea in Nuclear Matter}
\author{P.J. Ellis and S.B. Parendo \protect\\
{\small School of Physics and Astronomy, University of Minnesota,}
\protect\\{\small Minneapolis, MN 55455}
\protect\\M. Prakash \protect\\{\small Physics Department, SUNY at Stony
Brook,}
\protect\\{\small Stony Brook, NY 11794}}
\date{~}
\maketitle
\thispagestyle{empty}
\begin{abstract}
We have examined the contribution of the filled negative energy sea of hyperons
to the energy/particle in nuclear matter at the one and two loop levels.
While this has the potential to be significant, we find a
strong cancellation between the one and two loop contributions for our
chosen parameters so that hyperon effects can be justifiably neglected.
\end{abstract}
\vskip-18cm
\hfill NUC-MINN-95/20-T\\
\newpage

In the vacuuum Dirac particles have a filled negative energy
sea with the positive energy states empty. Zero-point fluctuations give rise to
the vacuum energy. In matter, the presence of positive energy particles
results in a modification of the vacuum energy due to interactions.
At the hadronic level, this implies that, even though only nucleon positive
energy states are filled in nuclear matter (provided the density is not too
high), the negative energy seas of particles other than nucleons  can play
a role. The purpose of this paper
to investigate the situation for the hyperons, which, together with the
nucleons, make up the basic octet. Since the hyperon coupling constants
are generally not well known, our aim is to obtain a first semi-quantitative
estimate to ascertain whether significant uncertainty is introduced into
the description of simple equilibrium nuclear matter.
We shall carry out our study for the one loop (Hartree) and two loop (Fock)
diagrams of Fig. 1, since, intuitively, one might expect some
cancellation between them, although as we shall see this is by no means
always the case.

We shall employ the Walecka Lagrangian, which has been
widely used in relativistic calculations of nuclear matter and finite nuclei
\cite{sew}. Here the baryonic interactions are generated by
the exchange  of scalar $\sigma$ and vector $\omega$ mesons.
The Lagrangian takes the form
\begin{eqnarray}
{\cal L} &=& \!\sum_B\bar B\left(i\gamma^{\mu}\partial_{\mu}-g_{\omega B}
\gamma^{\mu}\omega_{\mu}
-\!M_B+g_{\sigma B}\sigma\right)\!B \nonumber \\
&& -\tquar \omega_{\mu\nu}\omega^{\mu\nu}\!
+\thalf m^2_{\omega}\omega_{\mu}\omega^{\
mu} +\thalf\partial_{\mu}\sigma\partial^{\mu}\sigma
-\thalf m^2_{\sigma}\sigma^2 \;, \label{hyp1}
\end{eqnarray}
where $M_B$ is the vacuum baryon mass and the field strength tensor
is $\omega_{\mu\nu}=\partial_{\mu}\omega_{\nu}-\partial_{\nu}\omega_{\mu}$.
The baryons $B$ included in the sum are nucleons, $n,p$,
and hyperons, $\Lambda,\Sigma^+,\Sigma^-,\Sigma^0,\Xi^-$ and $\Xi^0$.

For purely nucleon degrees of freedom many calculations have been
carried out in the relativistic Hartree approximation which includes the
one loop diagram of Fig. 1(a). More recently the exchange or Fock
contribution, represented by the two loop diagram of Fig. 1(b), has been
examined. The first calculation was carried out by Furnstahl, Perry and
Serot \cite{2loop} and the diagram was found to be very large. Subsequently
it was pointed out that the composite nature of nucleons suggests that form
factors should be included at the vertices; these are represented by
the heavy dots in Fig. 1(b). Using  phenomenological monopole form factors
the two loop contribution was very substantially reduced \cite{ff} so that
it was much smaller than the one loop result. Recently progress has been
made in attempting to calculate the form factors by allowing for
the exchange of soft vector mesons at the vertex \cite{vertex}. The results
are quite similar to those obtained with phenomenological form factors.

Allowing for the presence of hyperons, the one loop Hartree
expression for the sigma field is
\begin{equation}
m_{\sigma}^2\sigma = \frac{2g_{\sigma N}}{\pi^2}
\int\limits_0^{k_{f}} dk\,k^2\frac{M^*_N}{\sqrt{k^2+M^{*2}_N}}
+\sum\limits_Bg_{\sigma B}\frac{\partial\Delta {\cal E}(M_B^*,M_B)}
{\partial M^*_B}
\;,\label{hyp5}
\end{equation}
where $N$ denotes a nucleon and the baryon effective masses are
$M^*_B=M_B-g_{\sigma B}\sigma$. The density is related to the Fermi momentum
by the usual expression $n=2k_f^3/(3\pi^2)$.
For the renormalization needed for the negative energy sea we use the
standard Chin-Walecka \cite{chin} prescription, so that
\begin{eqnarray}
\Delta {\cal E}(M^*,M)&\pj=&\pj-\frac{1}{8\pi^2}\biggl[
M^{*4}\ln\frac{M^*}{M}+M^3(M-M^*)-{\textstyle\frac{7}{2}}M^2(M-M^*)^2
\nonumber\\
&&\qquad\qquad\qquad+{\textstyle \frac{13}{3}}
M(M-M^*)^3-{\textstyle \frac{25}{12}}(M-M^*)^4\biggr]\;.
\label{walvac}
\end{eqnarray}
The values of the nucleon parameters  used are
$g_{\sigma N}^2= 54.3$, $m_\sigma = 458$ MeV, $g_{\omega N}^2 =
102.8$ and $m_\omega = 783$ MeV, which saturate nuclear matter (without
hyperons) at
$k_f=1.3$ fm$^{-1}$ with a binding energy of 15.75 MeV \cite{2loop}.
The contribution to the energy density from a given hyperon species at the
one loop level is then $\Delta E(M^*_H,M_H)$ for which we need to specify the
hyperon coupling constants.

The two-loop contributions we treat perturbatively since they are small
when form factors are employed \cite{ff,vertex}. Formally, using point
vertices Fig. 1(b) gives a contribution to the energy density
\begin{equation}
{\cal E}_{pt}= \frac{1}{2}\, \int \frac{d^4p_1}{(2\pi)^4} \frac{d^4p_2}{(2\pi)
^4}\,{\rm Tr}[S(p_1)\Gamma_1S(p_2)\Gamma_2]D(k) \, ,
\end{equation}
where  $S$ ($D$) represent baryon (meson) propagators, $\Gamma$ represents
a point vertex, $k=p_1-p_2$ and Lorentz indices have been suppressed.
For hyperons, the two loop contribution is due to vacuum fluctuations and
we use the expression given in Ref. \cite{2loop}. We allow for the
composite nature of hyperons by introducing form factors.
At each vertex we employ a phenomenological monopole form,
$(1 - q^2/\Lambda_{\rm cut}^2)^{-1}$, where $q$ is the four-momentum
transfer to the meson ($q^2<0$) and $\Lambda_{\rm cut}$ is a cut-off parameter;
alternatively this could be viewed as a single dipole form factor at one of
the vertices. The energy density can be written
\begin{equation}
{\cal E} = \frac {\Lambda^2_{\rm cut}}{\Lambda^2_{\rm cut} -m^2}
\bigg[ \frac {\Lambda_{\rm cut}^2}{\Lambda^2_{\rm cut} -m^2}
\bigg( {\cal E}_{pt} (m^2) - {\cal E}_{pt} (\Lambda_{\rm cut}^2) \bigg) +
\frac {d{\cal E}_{pt} (\Lambda_{\rm cut}^2)}{d\ln \Lambda^2_{\rm cut}}
\bigg] \, ,
\end{equation}
where $m$ is the mass of the exchanged meson.

Since the hyperon parameters which enter are poorly known, it is
well to begin with a simple exploration of parameter space. Thus we focus
on a single hypothetical hyperon and consider a vacuum mass of 1 or 2 GeV.
We choose the ratios of the hyperon to nucleon coupling constants,
$x_{\sigma H}=g_{\sigma H}/g_{\sigma N}$ and
$x_{\omega H}=g_{\omega H}/g_{\omega N}$, to both be $\twoth$, as suggested by
simple quark counting arguments.
The contribution to the energy/particle, {\it i.e.} ${\cal E}/n$, of nuclear
matter is shown in Table 1. Results are given for values of the nucleon
Fermi momentum of $k_f$=1.3, 1.6 and 1.9
fm$^{-1}$, corresponding to densities of $n_0$, 1.9$n_0$ and
3.1$n_0$, respectively, where equilibrium nuclear matter density is denoted
by $n_0$. These densities should be below the threshold for positive energy
hyperons. Consider first the cases where $\Lambda_{\rm cut}=\infty$,
{\it i.e.} the form factor is simply unity. The leading contribution to the
energy/particle for both the one and two loop results is proportional to
$(M^*_H-M_H)^5/M_Hk^3_f$. This accounts quite well for the relative
magnitudes of the results, although when comparing different hyperon masses
at the two loop level it must be borne in mind that there is a residual
dependence on the ratio of the meson mass to the hyperon mass which results
in a 12\% deviation from this simple formula. The negative two loop
contribution from $\omega$-exchange dominates so that the overall
contribution from the negative energy sea is negative, as with nucleons, but
the magnitude is smaller here since the effective mass is larger. When we
introduce form factors with $\Lambda_{\rm cut}=2$ GeV the dominant feature
is that the $\omega$ contribution is reduced by a larger amount than the
$\sigma$ contribution-- by a factor of $\sim2$ (5) for $M_H=1$ (2) GeV. This
is sufficient for the net two-loop result to be positive for $M_H=2$ GeV.
The case $\Lambda_{\rm cut}=1$ GeV
is more dramatic in that for $M_H=1$ GeV the two-loop $\sigma$ and $\omega$
contributions essentially cancel, while for $M_H=2$ GeV the $\omega$
contribution becomes positive. For a given $M_H$ and $\Lambda_{\rm cut}$
the relative values at different values of $k_f$ still follow the leading
order behavior mentioned above.
We see that the two loop results are sensitive to
the hyperon mass and give a net positive result for the larger mass,
assuming $\Lambda_{\rm cut}$ is the same for both mesons, and this would
slightly increase the one loop contribution.

We turn now to the effect of the filled negative energy sea of
$\Lambda,\ \Sigma$ and $\Xi$ hyperons upon the energy of nuclear matter.
Following Glendenning and Moszkowski \cite{glenmos}, we constrain the
coupling constants of the $\Lambda$-hyperon by requiring that the correct
energy be obtained for the lowest $\Lambda$ level in nuclear matter
at saturation. Defining, as before, the ratio
$x_{\sigma\Lambda}=g_{\sigma\Lambda/}g_{\sigma N}$ {\it etc.}, this gives at
the
Hartree level
\begin{equation}
-28=x_{\omega\Lambda} g_{\omega N}\omega_0-x_{\sigma\Lambda}
g_{\sigma N}\sigma\;,
\label{hyp6}
\end{equation}
in units of MeV. Choosing $x_{\omega\Lambda}=\twoth$ on the basis of quark
counting arguments, we find $x_{\sigma\Lambda}=0.614$ and these values
yield reasonable properties for neutron stars according to Ref. \cite{glenmos}.
For the $\Sigma$ we rely on a recent analysis of $\Sigma^-$ atomic data
by Mare\v{s} et al~\cite{sigmacou}. For consistency we pick their case with
$x_{\omega\Sigma}=\twoth$ which yields a good fit
with $x_{\sigma\Sigma}=0.54$. As regards the
$\Xi$, in the absence of better information we take the simple quark model
estimate $x_{\omega\Xi}=x_{\sigma\Xi}=\oneth$. Using the analogue of
Eq. (\ref{hyp6}), the energy of the lowest $\Sigma$ and $\Xi$
levels in equilibrium  nuclear matter are $-10$ and $-21$ MeV respectively.
The latter is reasonable according to
the discussion of Schaffner et al. \cite{djm}.  As regards the $\Sigma$
energy, Mare\v{s} et al~\cite{sigmacou} suggest that the optical potential
is repulsive in the nuclear interior. A modest adjustment of
$x_{\sigma\Sigma}$ to yield zero or a small positive energy would
not qualitatively change our results. Finally we need to specify
the form factors. The J\"ulich fit to hyperon-nucleon scattering \cite{hypn}
indicates that, unlike nucleons, hyperons favor a larger cut-off for
$\omega$ exchange, $\Lambda_{\rm cut}=2$ GeV, than for $\sigma$ exchange,
$\Lambda_{\rm cut}=1$ GeV. The cut-off there was in the 3-momentum, but we
shall simply adopt these values as reasonable estimates.

The effect of the hyperons upon the self-consistent $\sigma$ field
equation  (\ref{hyp5})
is slight. It results in an $\sim1$\% reduction in the field which does
not significantly affect the nuclear matter saturation properties obtained
with nucleons alone. The hyperon contributions to the energy/particle,
with the parameters discussed above, are shown in
Table 2. The $\Xi$ is less important than the other hyperons due to its
smaller coupling and resulting larger effective mass. The total hyperon
one-loop contribution amounts to 3--4\% of the pure nucleon
energy/particle at $k_f=1.3$ and 1.9. This is modest, but not negligible.
The two-loop contribution is dominated by $\omega$ exchange so that for
each hyperon species the net result is negative. At all three values of
$k_f$, the total hyperon results at the one and two loop levels show a
strong cancellation so that the net hyperon
contribution is negligible. Note that this depends on using a larger
$\Lambda_{\rm cut}$ for the $\omega$ which gives a sizeable, negative
two-loop result ({\it cf.} Table 1). We should also comment briefly on
uncertainties in the coupling constants. Consider the $\Sigma$:
Mare\v{s} et al~\cite{sigmacou} report that their best fits are obtained
with  $x_{\omega\Sigma}$ in the range $\twoth$ to 1. With
$x_{\omega\Sigma}=1$ they obtain $x_{\sigma\Sigma}=0.77$ which, with our
parameters,  would give the lowest $\Sigma$ level in nuclear matter
at $-4$ MeV. Since the sigma coupling is larger here the effective mass is
smaller which leads to enhanced contributions, for example at saturation
the one plus two loop contribution
is $1.6-3.0=-1.4$ MeV,  by no means negligible.
Also the J\"ulich group \cite{hypn} have suggested a large value of
$x_{\sigma\Sigma}=1.28$, while $x_{\omega\Sigma}$ remains $\twoth$.
This yields a large value for
the net one plus two loop contribution of 14.5 MeV, but this should not be
taken seriously since these parameters yield an unreasonably large binding
for the lowest $\Sigma$ level in nuclear matter.

Our aim has been to investigate the contribution of the filled negative
energy sea of hyperons to the binding energy of nuclear matter. With
the most reasonable parameters that we could find there was a strong
degree of cancellation between the one and two loop contributions so that
the sum was negligibly small. This is very satisfactory since it implies
that such effects can be safely neglected, as has been common practice.
This conclusion is, however,
strongly dependent on the rather uncertain values of the parameters
and further efforts to tie them down would be very valuable.

This work was supported in part by the U. S. Department of Energy
under grant numbers DE-FG02-87ER40328 and DE-FG02-88ER40388.

\newpage
\small
\centerline{{\bf Table 1.} Contribution of a single hypothetical hyperon, of
mass 1 or 2 GeV, to the}
\centerline{energy/particle at the one and two loop levels. Results are
given for several values}
\centerline{of the Fermi momentum, $k_f$, and the form factor
cut-off parameter, $\Lambda_{\rm cut}$.}
\begin{center}
\begin{tabular}{|ccc|c|crrr|}
\hline \hline
&&&&\multicolumn{4}{|c|}{Two Loop}\\
$k_f$ & Mass $M_H$ & $M^*_H/M_H$ & One Loop & Cut-off $\Lambda_{\rm cut}$&
$\sigma$ & $\omega$ & Total\\
fm$^{-1}$ & GeV & & MeV & GeV &MeV & MeV & MeV\\ \hline
    &   &      &       & $\infty$ & 0.397 & $-1.555$ & $-1.158$\\
    & 1 & 0.83 & 0.307 & 2        & 0.199 & $-0.393$ & $-0.194$\\
    &   &      &       & 1        & 0.068 & $-0.067$ & $0.001$ \\
1.3 & & & & & & & \\[-5mm]
    &   &      &       & $\infty$ & 0.227 & $-0.881$ & $-0.655$ \\
    & 2 & 0.92 & 0.153 & 2        & 0.041 & $-0.030$ & 0.010 \\
    &   &      &       & 1        & 0.008 & 0.005    & 0.013 \\ \hline
    &   &      &       & $\infty$ & 1.335 & $-5.234$ & $-3.899$ \\
    & 1 & 0.76 & 1.058 & 2        & 0.682 & $-1.360$ & $-0.678$\\
    &   &      &       & 1        & 0.235 & $-0.233$ & 0.001 \\
1.6 & & & & & & & \\[-5mm]
    &   &      &       & $\infty$ & 0.775 & $-3.013$ & $-2.239$ \\
    & 2 & 0.88 & 0.529 & 2        & 0.141 & $-0.105$ & 0.036 \\
    &   &      &       & 1        & 0.029 & 0.016    & 0.045 \\ \hline
    &   &      &       & $\infty$ & 2.307 & $-9.047$ & $-6.741$\\
    & 1 & 0.70 & 1.866 & 2        & 1.198 & $-2.405$ & $-1.207$\\
    &   &      &       & 1        & 0.413 & $-0.415$ & $-0.001$\\
1.9 & & & & & & & \\[-5mm]
    &   &      &       & $\infty$ & 1.352 & $-5.260$ & $-3.908$ \\
    & 2 & 0.85 & 0.930 & 2        & 0.248 & $-0.185$ & 0.062 \\
    &   &      &       & 1        & 0.050 & 0.029    & 0.079 \\ \hline\hline
\end{tabular}
\end{center}
\newpage
\normalsize
\begin{quote}
{\bf Table 2.} Contribution of the $\Lambda,\ \Sigma$ and $\Xi$ hyperons
to the energy/particle at the one and two loop levels
for three values of the Fermi momentun, $k_f$. The parameterization is
discussed in the text.
\end{quote}
\begin{center}
\begin{tabular}{|ccc|c|ccc|}
\hline \hline
&&&&\multicolumn{3}{|c|}{Two Loop}\\
$k_f$ & Particle & $M^*_B/M_B$ & One Loop & $\sigma$ & $\omega$ & Total\\
fm$^{-1}$ & & & MeV &MeV & MeV & MeV\\ \hline
    & $\Lambda$  & 0.86  & 0.179 & 0.028 & $-0.188$ & $-0.160$\\
1.3 & $\Sigma$   & 0.89  & 0.273 & 0.030 & $-0.251$ & $-0.221$\\
    & $\Xi$      & 0.94  & 0.014 & 0.000 & $-0.003$ & $-0.002$ \\ \hline
Totals &         &       & 0.466 &       &          & $-0.383$ \\ \hline
    & $\Lambda$  & 0.80  & 0.609 & 0.096 & $-0.643$ & $-0.547$\\
1.6 & $\Sigma$   & 0.83  & 0.927 & 0.103 & $-0.856$ & $-0.754$\\
    & $\Xi$      & 0.91  & 0.048 & 0.002 & $-0.009$ & $-0.007$ \\ \hline
Totals &         &       & 1.584 &       &          & $-1.308$ \\ \hline
    & $\Lambda$  & 0.75  & 1.068 & 0.169 & $-1.132$ & $-0.963$\\
1.9 & $\Sigma$   & 0.79  & 1.623 & 0.180 & $-1.505$ & $-1.325$\\
    & $\Xi$      & 0.89  & 0.083 & 0.003 & $-0.015$ & $-0.013$ \\ \hline
Totals &    &       & 2.774 &       &          & $-2.301$ \\ \hline\hline
\end{tabular}
\end{center}
\newpage
\centerline{{\bf Figure Caption}}

\noindent Fig. 1. (a) One loop Hartree and (b) two loop Fock diagram; the
full (dashed) lines represent baryons (mesons) and the heavy dots indicate
form factors.

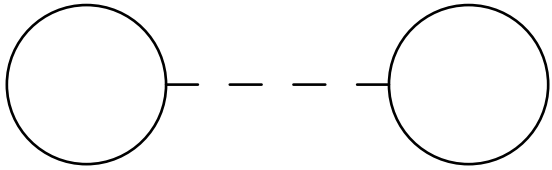
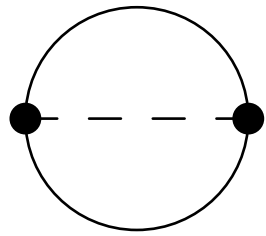
\end{document}